# Using Photo Modeling Based 3DGRSL to Promote the Sustainability of Geo-Education, a case study from China.

**Xuejia Sang**[1], **Linfu Xue**[2], **Xiaopeng Leng**[3], **Xiaoshun Li**[1,*], **Jianping Zhou**[4]

[1]   School of Environment Science and Spatial Informatics, China University of Mining and Technology, Xuzhou 221116, China; sangxj@cumt.edu.cn (X.S.), lixiaoshun1983@sohu.com (X.L.)

[2]   College of Earth Science, Jilin University, Changchun 130061, China; xuelf@jlu.edu.cn (X.L.)

[3]   College of Information Sciences and Technology, Chengdu University of Technology, Chengdu 610059, China

[4]   University of Göttingen, Geoscience Center, Department of Sedimentology and Environmental Geology, Goldschmidtstrasse 3, D-37077, Germany; jzhou1@uni-goettingen.de (Z.J.)

*   Correspondence: lixiaoshun1983@sohu.com; Tel.: +86-132-580-03887

**Abstract:** In earth science education, observation of field geological phenomena is very important. Due to China's huge student population, it is difficult to guarantee education fairness and teaching quality in field teaching. Specimens are indispensable geo-education resources. However, the specimen cabinet or picture specimen library has many limitations and it is difficult to meet the internet-spirit or geo-teaching needs. Based on photo modeling, this research builds a 3D Geo-Resource Sharing Library (3DGRSL) for Geo-Education. It uses the Cesium engine and data-oriented distributed architecture to provide the educational resources to many universities. With Browser/Server (B/S) architecture, the system can realize multi-terminal and multi-scenario access of mobile phones, tablets, VR, PC, indoor, outdoor, field, providing a flexible and convenient way for preserving and sharing scientific information about geo-resources. This makes sense to students who cannot accept field teaching in under-funded colleges, and the ones with mobility problems. Tests and scoring results show that 3DGRSL is a suitable solution for displaying and sharing geological specimens. Which is of great significance for the sustainable use and protection of geoscience teaching resources, the maintenance of the right to fair education, and the construction of virtual simulation solutions in the future.

**Keywords:** Geo-Resource Sharing Library; Photographic modeling; Virtual Lab

## 0. Introduction

In Earth science, many field educational resources are so important rare and non-transportable resources that many universities offer rich field practice courses [1-2]. Examples include the Colorado School of Mines, New York University, and the University of Alberta [3-5]. The British Geological Society recommends that a minimum field practice period for undergraduate education in geology as 105 days [6]. However, China's special problems have made most geo-practical courses time-critical and under-quality. How to resolve the conflict between quantity and quality has always been a difficult problem for field geo-education. Virtual visualization looks like a solution. However, modeling of real rocks has always been very time-consuming and laborious, and rough and false virtual rocks, outcrops, and terrain have severely restricted their application in geosciences. Therefore, the reproduction technology of real field environment is the core issue of virtual geoscience research, and it is also an important issue for the open and sharing of geo-education resources to promote the sustainability and equity of geoscience education.

## 1. Problems of Geo-Educational Resources in China

*1.1 Restriction for geo-education*

China has the largest group of young students in the world. It is difficult to manage field activities and safety issues are frequent. The geo-educational resources are often located in inaccessible terrains such as steep cliffs and deep valleys. Even with thoughtful considerations, field-teaching activities are highly risky. In 2018, a tutor and four students from Southwest Petroleum University and Jilin University died during field geological survey; a graduate student of China University of Geosciences fell off a cliff in 2015.

Apart from the safety issues, summer field practices are often delayed or cancel by continuous rain in Eastern China. There are many more restrictions, just like Corona Virus Disease 2019, which has stopped classroom teaching and field teaching activities in almost all Chinese universities since the Chinese New Year. Field activities are facing so many restrictions that many schools have begun to reduce field courses, and some underfunded schools have directly canceled them. This has seriously affected the quality of geo-education in China [3,6].

*1.2. Lack of protection for geo-education resources*

The real-world teaching resources often face the risk of being destroyed. For example, many geological samples have been destroyed during civil construction. Many non-renewable fossils, minerals, and crystals have been destroyed or transferred to traditional specimen rooms (Figure 3).

There is a geological practice base in the Jilin University in Xingcheng City. Compared with the other seven practice bases in China, the Xingcheng base has better infrastructure and classic strata and structural outcrops. However, in 2018, to prevent landslides and falling rocks (Figure 1), the local government protected the exposed rock mass on the roadside, thus it is no longer available for teaching. In other words, this classic field geo-teaching resource has lost.

In another example, in Linyi City, China, footprint fossils were reported by Xing [7], evidencing that the Deinonychosauria lived in groups. However, these were destroyed by thieves in less than a year, leaving only gravel on the ground (Figure 2).

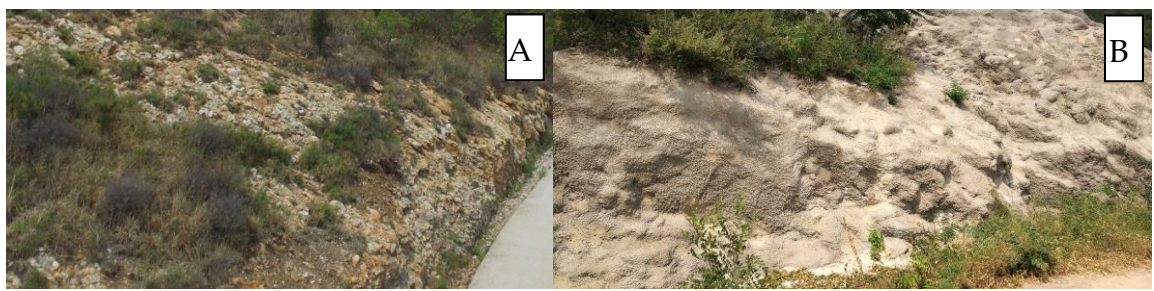

**Figure 1.** A geological site that was "damaged" while protecting the safety of citizens. **A)** Sandstone outcrop in Xingcheng City, China, photo taken on June 11, 2017. **B)** The outcrop was completely covered by cement, photo taken on September 11, 2018.

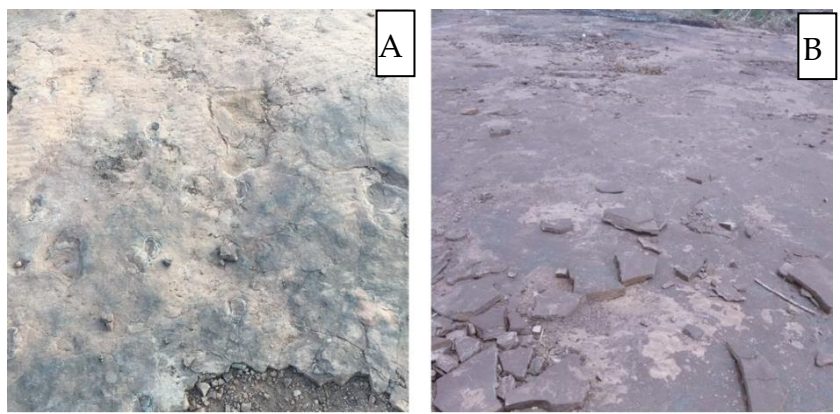

**Figure 2.** Geological sites in nature are not effectively protected. **A)** Dinosaur footprint fossil in Linyi City, China, photo taken on June 1, 2018. **B)** Dinosaur footprint fossils were destroyed, photo taken on April 26, 2019.

### 1.3. The inconvenience of specimens storing

First, there are huge specimens that cannot be collected in nature, such as mountains, rivers, valleys, cliffs, and boulders. They are important rare and non-transportable resources for geo-education. Second, the collection and sharing of even a small geoscience specimen are difficult. Classic specimens are often considered an exclusive collection. At the same time, many specimens were locked in the cabinet, making it difficult to observe freely (Figure 3). Moreover, many under-funded geoscience colleges are unable to provide students with appropriate geo-specimens or practical courses. Therefore, preservation and utilization of these specimens have always been a problem.

**Figure 3.** The traditional method for storing geo-specimens **A)** A dusty specimen display case. **B)** A locked specimen display case to prevent theft but it also prevents the students from observing it.

### 1.4. Inappropriateness of Geo-image database

In order to facilitate the view and digitization of specimens, many specimen databases have been created, such as the National Specimen Information Infrastructure (NSII) [8], National Infrastructure of Mineral, Rock and Fossil Resources for Science and Technology (NIMRFRST) [9]. However, most of these databases store the data in the form of images and texts (Figure 4). These databases merely offer directory or indexing of resources. These images and texts are often boring to students. A user cannot see the three-dimensional shape of the specimen, the spatial shape before the collection, and the macroscopic relationship between the specimen and the environment. Therefore, these databases are not conducive to the user to develop a whole way of geo-thinking.

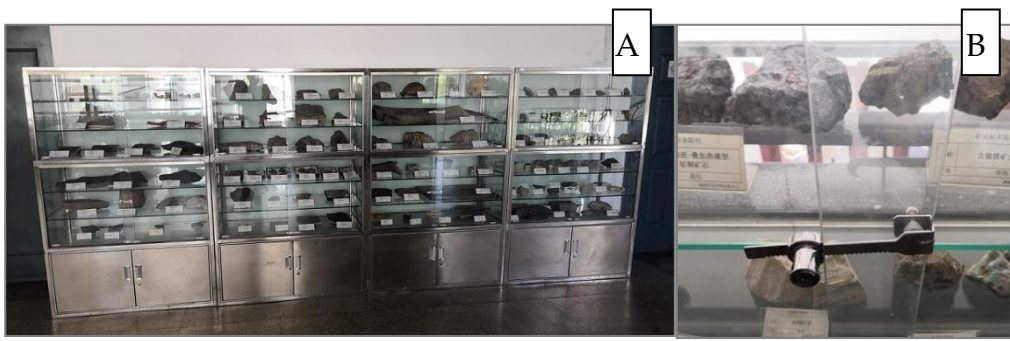

**Figure 4.** A typical Image geo-database uses pictures and text to describe specimens. **A)** Typical online specimen picture database named NSII. **B)** Database index for NIMRFRST.

## 2. Original Intention of 3DGRSL

Skibba [10], Jones [11] and Owens [12] believe that 3D and VR Technologies can be a solution to STEM education like geology, geomorphology or topography [13]. Carbonell found that the aid of 3D visualization can greatly improve the teaching in geoscience [14, 15]. Meyer believes that virtual reality technology allows students to explore the earth while sitting in a classroom [16]. The virtual lab has been proven to be the right way to promote effective teaching. However, the teaching of geosciences needs special attention. Since the knowledge of geology comes from nature, the virtual environment used in geology education must be realistic. Geology, geomorphology, and geography education require real surface-morphology relationships; structural analysis requires a rigorous and accurate surface model; lithology analysis requires detailed model texture. These natural phenomena should not be fabricated. Because, after all, the ability to interpret natural phenomena is to be verified in the actual work of students in the future.

The game map editor (GME) is not the right tool for creating virtual environments directly [11]. Traditional workflow with GME usually depends on the Digital Elevation Model (DEM) and use satellite image as a ground-texture. The virtual terrain and the textural fineness are limited by the resolution of the DEM and the satellite image, respectively. Unfortunately, this method cannot meet the observing needs of geoscience at the outcrop and specimen scale. Figure 5 gives the process of producing terrain. First, the DEM data (Figure 5**B**) is used to stretch out the 3D surface, and then the satellite image texture (Figure 5**A**) is attached to the 3D surface. However, the resolution of the satellite image is too low and can only be manually replenished with texture information. Figure 5**C** is the final model and Figure 5**D** is the close-up view of the model. In Figure 5**D**, it can be observed that the terrain lacks details and the texture was repeated and false. Limited to GME, many virtual geosciences teaching systems cannot provide real terrain and rock models (Figure 6) [17].

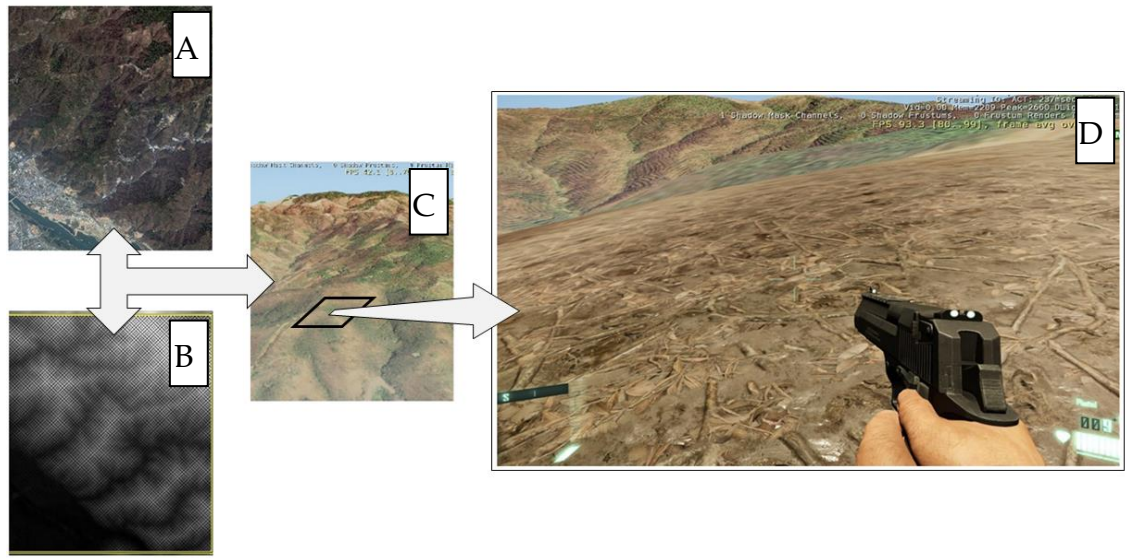

**Figure 5.** The traditional work process of establishing a ground-model based on DEM data and satellite images. **A)** Satellite image. **B)** Digital Elevation Model (DEM). **C)** Ground-model made by game map editor (GME). **D)** Detail of ground-model.

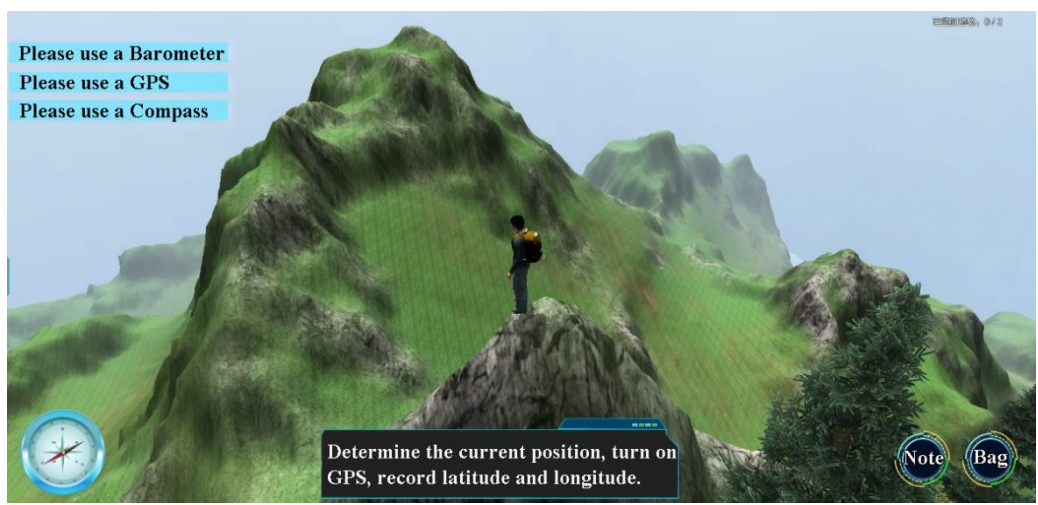

**Figure 6.** Fake terrain and outcrops commonly used in a virtual education system.

Fake terrain and stretched texture are not problematic in video games, but they cause a severe problem in geoscience education or libraries. The real models based on true geoscience knowledge are critical for both digital archiving and education. Therefore, the difficulty of modeling complex and realistic 3D models is an important issue that needs to be solved during the construction of 3D Geo-Resource Sharing Library.

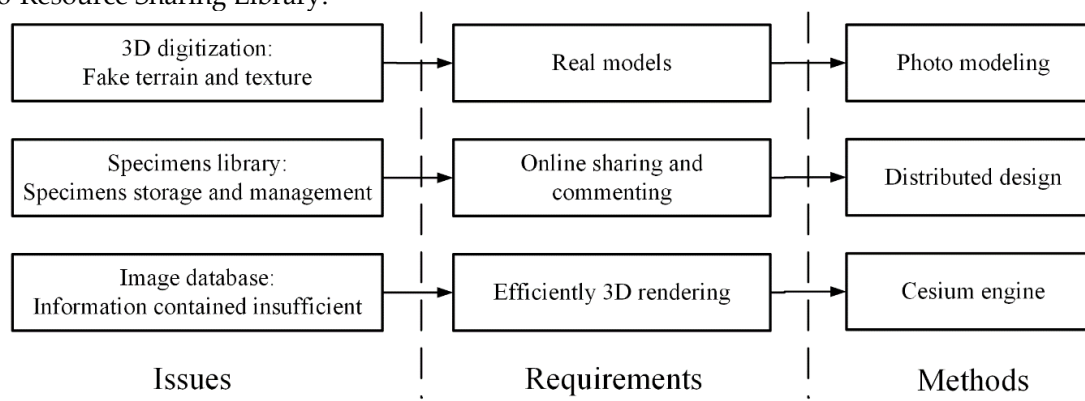

**Figure 7.** Design ideas for some features of 3DGRSL.

Therefore, this study puts forward the idea of building 3DGRSL (Figure 7) which can be used to show realistic geoscience 3D models. The ideal 3DGRSL should have some obvious advantages:

1. The real three-dimensional model can be directly observed in geoscience.

2. Shared and open design, which is convenient for researchers to maintain, update and communicate.

3. Efficient 3D display mode to meet the needs of 3D display efficiency and hardware.

In view of the above, new methods need to be applied. Next, we introduce several methods for targeted problem solving in 3DGRSL.

## 3. Methodology

### 3.1. Photo modeling

The photo modeling method with core algorithms such as structure from motion (SFM) and multi-view stereo (MVS) [18] provides high-precision 3D data comparable to laser scanning while offering portability, low power consumption, and fast data collection. Because of these advantages, photo modeling technology has been applied rapidly in geosciences in the past four years [18-23], and there is a trend to replace ground laser scanning modeling. They are considered important methods for solving problem about producing complex 3D model [22, 24].

The first step in photo modeling is the feature matching of the SFM algorithm (Figure 8). The SFM looks for the corresponding features of images in a set and uses the resulting matching points to calculate the relative position of each camera. It simultaneously determines the camera position at the time of each image capture and the spatial coordinates of each matching feature point [18].

This algorithm typically uses an incremental approach that combines more images in each successive iteration. These matching feature points constitute a feature point cloud that represents the structure of the target surface defined within the local coordinate system. After obtaining the attitude information of the known camera, a multi-view stereo algorithm is used to generate a dense point cloud for the surface of the object (Figure 9).

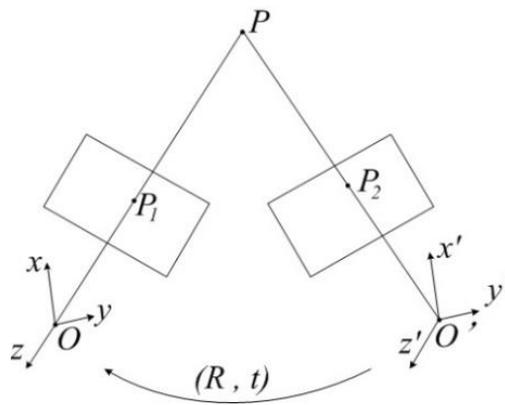

**Figure 8.** Schematic diagram of Structure from Motion (SFM) algorithm.

The MVS algorithm identifies the best match between the images by systematic search on the pixel grid. It improves the matching precision and obtains more 3D point clouds known as dense point clouds [19]. There is no technical problem in using triangulation or mesh interpolation to generate a digital elevation model (DEM) or a 3D model of any selected direction from these dense point clouds. The 3D model or DEM can be used to correct the ground undulation and derive the orthophoto [23]. In addition, the 3D model can be formatted into a common 3D format, like 3D tiles for visualization and sharing [25].

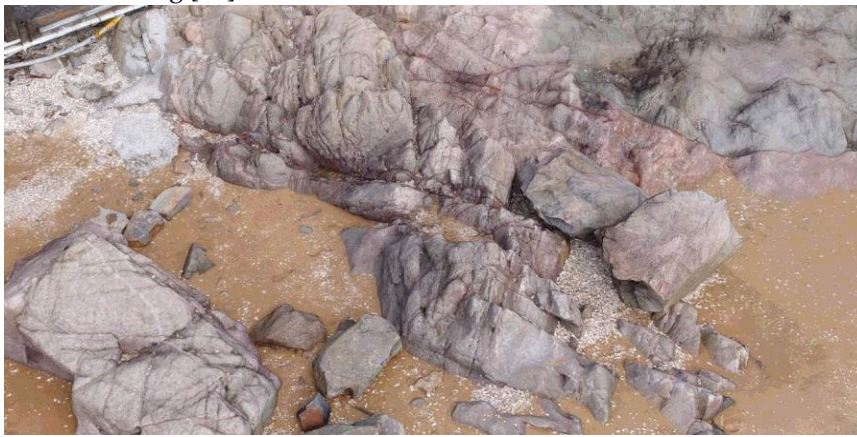

**Figure 9.** Using SFM to restore 3D form with high resolution.

### 3.2. Cesium Engine

Cesium Engine is an open-source JavaScript library that has been widely used in world-class 3D rendering and map hosting solutions. It creates maps or 3D globes based on static and dynamic geographic information data. Cesium has advanced 3DTiles technology that can improve the transmission and display efficiency of 3D model resources. The 3D Tiles format is a common 3D model transmission and presentation solution for platforms such as Cesium. Its tile-set is a collection of tiles organized by a tree-shaped spatial data structure. Each tile has a virtual box that contains its inner structure completely. Its structure has spatial coherence, and the content of the sub-tile is entirely contained within the virtual range of the parent tile giving it the flexibility of hierarchical, sub-regional, and multi-threaded loading, greatly improving the speed and efficiency of the model display (Figure 10). In February 2019, 3D Tiles became the OGC Community Standard for Streaming Massive 3D Geospatial Content [29].

The complicated 3D model generated by photographic modeling needs to be simplified in order to reduce resource expenditure as it has excessive detail and unnecessary illumination information even with 3DTiles. The Instant Mesh software or Simplygon can be used to simplify the mesh and then 3DsMax can be used to clean up the redundant vertices [26, 27]. According to previous experiments, the model can be reduced by 70% of surfaces without any loss of visual effects.

The models produced by photo modeling come with light and shadows information and must be removed. Black curtains, shadowless lamps, etc., can be used to eliminate shadow in the modeling process of smaller specimens. However, there are no shadowless-lights that can cover the cliff. The delighting tool can be used to remove shadows after modeling. It developed by Unity French team for generating the final unlit diffuse texture, and it is open-sourced on Github [28].

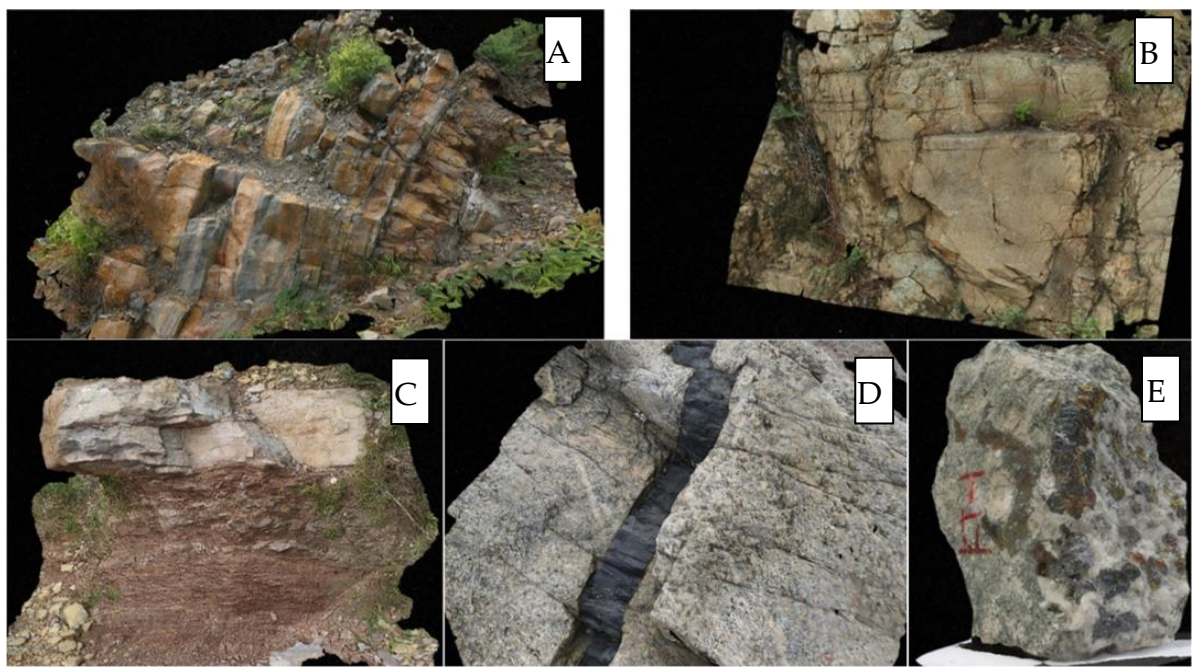

**Figure 10**. Fine-scale 3D specimens. A) Maroon feldspar sandstone B) Granite syenite C) Conformable contact between limestone and fuchsia shale D) Diabase vein invading granitic gneiss. E) Chalcopyrite.

*3.3. Distributed construction*

Distributed design is one of the characteristics of 3DGRSL. Several distributed data nodes support the platform data, the existing and future added data are uniformly registered then managed by the Data Registration Center. "Data" was separated from "functions" from the perspective of design and deployment. This not only facilitates data sharing and independent management of all units but also facilitates the development of "data-oriented" software and systems in the future, saving development costs and time. The platform is initially established with three service nodes, including China University of Mining and Technology, Jilin University, and Chengdu University of Technology, and the unique geo-resources of each unit are integrated (Figure 11).

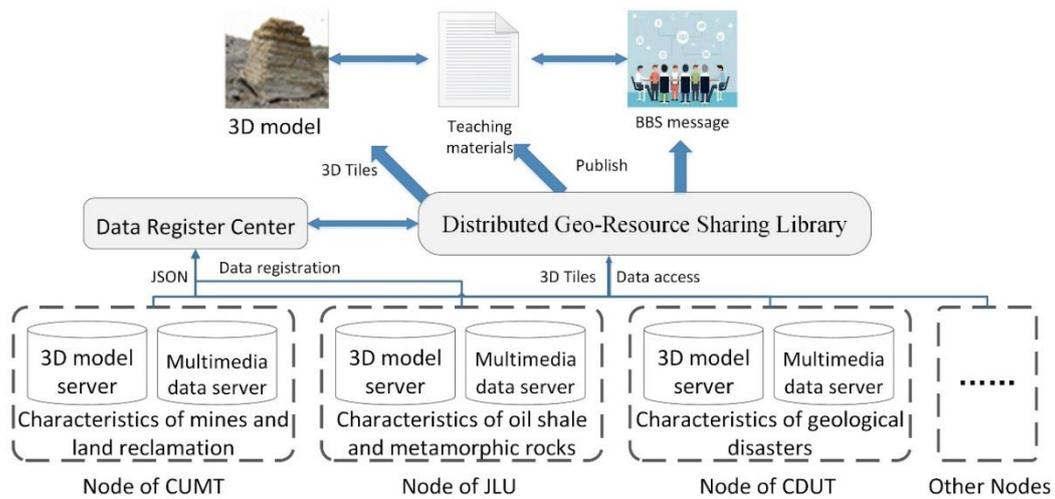

**Figure 11.** Schematic diagram of the 3DGRSL.

Basemap of 3DGRSL like terrain, vectors, and annotated web map tile service (WMTS) provided by the Tianditu Server [30]. Additional drones and aerial data are accessed in areas with severe cloud coverage. With GeoServer, other vector geological map data can also be added as required (Figure 11). The system can view and extract the attributes of a geological map to support the understanding of the regional geological background.

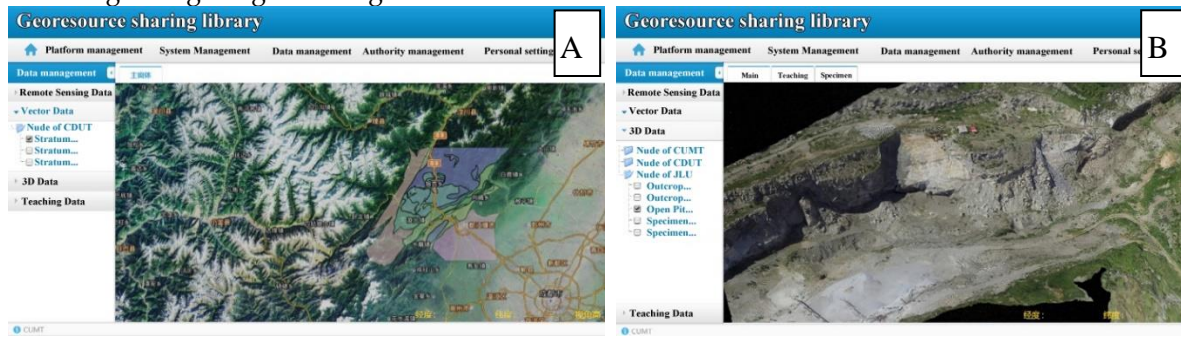

**Figure 12.** A system producing the geological background of a target area by displaying vector geological maps (A) or large-scale 3D models (B).

The model with a high-resolution map can clearly distinguish the gravel and shell on the surface; in the three-dimensional shape, it can be clearly seen that the vein is cut by fault (Figure 12). Users can learn about spatial distributions, scales, contact relationships, texture, and other characteristics of the field geology phenomena online.

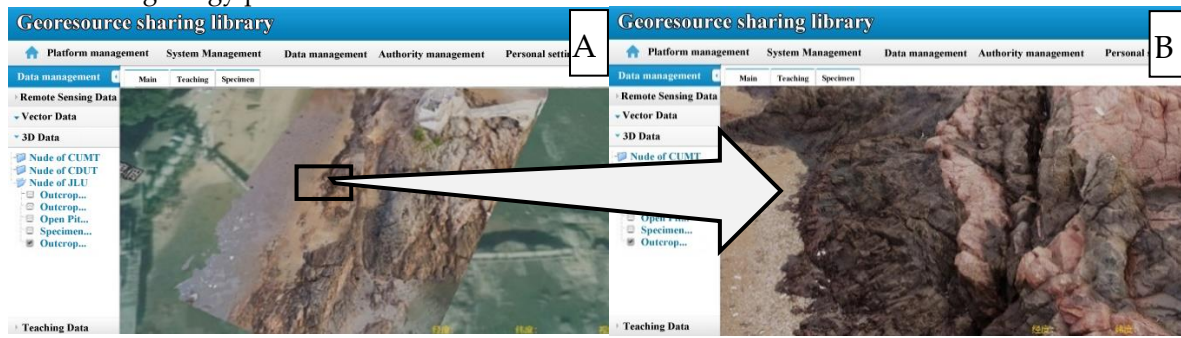

Figure 12. Outcrop-scale 3D models (A) and a vein in it (B).

In addition to the digital geological phenomena, the teaching resources in the form of graphic multimedia must also be provided. The system provides a multi-tag view for users to switch between the phenomenon tag and the data tag. The teaching materials are collected and placed around the typical phenomena of the target, with pictures and texts.

## 4. Evaluation

Twenty teachers and ten students with field geology experience (after obtaining informed consent) were selected to use the Geo-Resource Sharing Library and score the system using a questionnaire. The scoring was mainly based on the following criteria:

1. Ease of use: Users can easily browse through the functions of the virtual system.

2. Comfort: Users can use and learn in a system comfortably.

3. Interactivity: it indicates the possibility of interaction.

4. Visual effects: it indicates the degree of satisfaction of the user's Geo-observation.

5. Fluency: it indicates the performance requirements of the device when the system is running.

6. Sharing convenience: it indicates the ease of sharing data or physical objects.

7. Geoinformation content: it indicates the amount of original information carried by the data or the physical object.

8. Storage cost: it indicates the cost of storage of data or physical objects.

**Table 1.** Questionnaire score and description

| Score | Description |
|-------|-------------|
| 1 | Very inconvenient |
| 2 | Some difficulties, but they can be overcome. |
| 3 | The advantages and disadvantages are equally obvious |
| 4 | Although inconvenient, it still meets the requirements of geosciences teaching |
| 5 | Very suitable for geoscience teaching |

**Table 2.** Reliability statistics and scale statistics

| Method | Cronbach's α | Cronbach's Alpha Based on Standardized Items | Mean | Variance | Std. Deviation |
|--------|-------------|----------------------------------------------|------|----------|----------------|
| 3D | .898 | .909 | 36.70 | 10.355 | 3.218 |
| Pic | .630 | .639 | 33.90 | 4.990 | 2.234 |
| Spec | .714 | .703 | 26.6667 | 3.402 | 1.84453 |

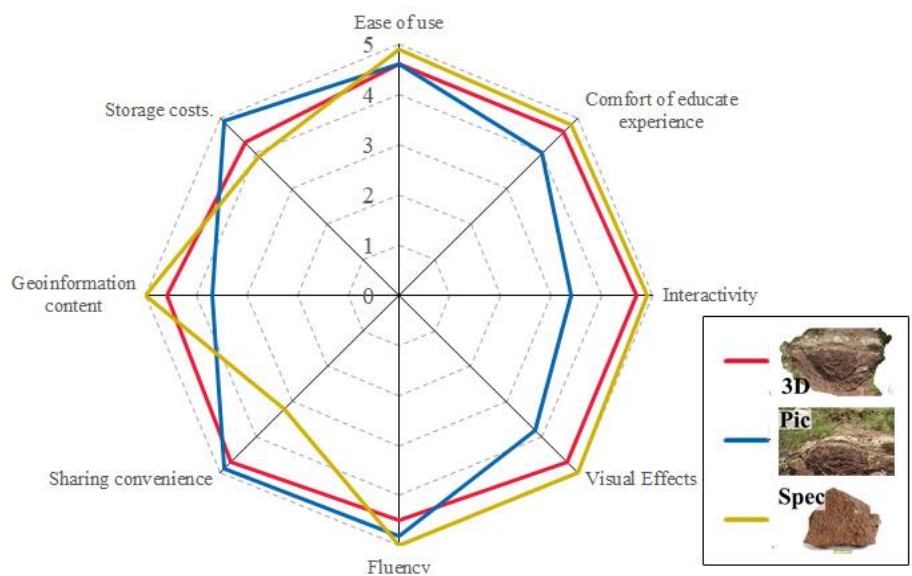

**Figure 14.** Comparison of three geo-database scores, including 3D model, picture, and specimen storage.

The specimen method is the most mature method, but its shortcomings include difficulty in sharing and storage. It can also be seen from Table 2 that the variance of the Spec score is small, indicating that most opinions tend to be concentrated. Cronbach's Alpha of Pic and Spec have a large score ($\alpha = 0.639$ & $0.703$). It seems that due to the different perceptions of the advantages by the raters. The scores of the 3D method seem largely, this may indicates that users are still skeptical about new technologies, and this situation needs time to improve.

As shown in Figure 14, physical specimens impede academic exchange, and specimens require large space for storage. Collecting information in the form of photographs is the most convenient method. The data size is small and transmission is easy. However, this method loses a lot of original information, and is not suitable for the specimen library.

## 5. Discussion

Solving field teaching difficulties and sharing geological specimens online are very important to improve the quality of geoscience teaching and promote the right to a fair education. How to tap the potential of digital resources for field geoscience phenomena and reduce the cost of repeated data collection is a problem that must be faced in the future construction of geoscience virtual teaching laboratories and simulation environments. The resource sharing library provides a good solution to this problem in an open, co-constructed, distributed and shared manner. Working with more geological colleges to share and digitize the characteristic geo-resources for geo-education is forward-looking for the future. This will not only preserve the geo-phenomena but will also allow students and researchers from all over the world to observe the phenomena and exchange ideas in the virtual geo-space.

Since the establishment of geosciences, the understanding of geology has been limited by their field-work experience. However, the idea of 3DGRSL may change the research and communication methods of the geoscience. People can tell a geological story by sending a URL instead of a bunch of static pictures or tens of pounds of stones.

With the development of the network and storage media, the data size will no longer be a problem. People will pay more attention to user experiences such as visualization effects and convenience. With Browser/Server (B/S) architecture, the system can realize multi-terminal and multi-scenario access of mobile phones, tablets, VR, PC, indoor, outdoor, field, providing a flexible and convenient way for preserving and sharing scientific information about geo-resources [31-33].

Taking geography education as an example, more intuitive and realistic display effects often lead to better education quality, which has been proven by many researchers [10-16]. We believe that this system can provide important ideas that construct new digital specimen libraries, improves the learning quality in the uneven education area, and guarantees a fair and sustainable development of geoscience education.

## 6. Conclusion

In this work, a new type of geo-specimen library, photo modeling based 3D Geo-Resource Sharing Library (3DGRSL) was built. From the statistical results, 3DGRSL is more intuitive and can render richer information than the image database. 3DGRSL has a display effect similar to the actual specimen, which greatly reduces the specimen storage cost and makes the sharing of specimens more convenient.

3DGRSL provides a solution of sharing to the characteristic geo-education resources for many universities. Resource integration of the three universities has been achieved in this study. This experiment proved the importance of photo modeling for online specimen libraries, especially geological specimens. The platform provides important and open data support for the sustainable use and protection of geoscience teaching resources, the maintenance of the right to fair education, and the construction of virtual simulation solutions in the future.

**Acknowledgments:** Thanks to Jianxiong Ma and other members of Digital Geoscience Research Center, College of Earth Sciences, Jilin University.

**Author Contributions:** Resources, Linfu Xue; Funding acquisition, Xiaoshun Li; Software, Xiaopeng Leng; Writing – original draft, Xuejia Sang; Writing – review & editing, Jianping Zhou. All authors have read and agreed to the published version of the manuscript

**Funding:** The Project supported by "the Fundamental Research Funds for the Central Universities" (No: 2020QN16).